\documentclass[preprint,prb,aps]{revtex4}
\usepackage{graphicx}
\usepackage{dcolumn}
\usepackage{bm}
\begin{document}
\title {Electron-phonon relaxation and excited electron distribution in
        zinc oxide  and anatase}
\author{ V.P. Zhukov $^{1,2}$}
\author{V.G. Tyuterev $^{3}$}
\author{E.V. Chulkov$^{2,4}$}
%
%
\affiliation{
   $^1$ Institute  of Solid State Chemistry,  Urals Branch of
       the Russian Academy of  Sciences,  Pervomayskaya 91,  620990,
        Yekaterinburg, Russia \\
   $^2$Donostia International Physics Center (DIPC), P.
       de Manuel Lardizabal, 4,
       20018, San Sebasti{\'a}n, Basque Country, Spain\\
   $^3$ Tomsk State Pedagogical University, 634041, Tomsk, Russia \\
   $^4$Departamento de F{\'\i}sica de Materiales, Facultad de
      Ciencias Qu{\'\i}micas, UPV/EHU and Centro de Fisica de Materiales CFM-MPC
      and Centro Mixto CSIC-UPV/EHU,
      Apdo. 1072, 20080 San Sebasti\'an, Basque Country, Spain \\
   }
\date{\today}
\date{\today}
\begin{abstract}
   We  propose a first-principle method for evaluations of  the time-dependent electron distribution function of excited  electrons in the
   conduction  band of semiconductors.    The  method takes into account the  excitations of electrons by external source  and
   the relaxation   to the bottom of conduction band via  electron-phonon coupling.  The  methods permits    calculations of   the
   non-equilibrium electron distribution function, the quasi-stationary
   distribution function with  steady-in-time source of light, the time of setting of the quasi-stationary  distribution and the time of
   energy loss  via relaxation  to the bottom of conduction band. The  actual  calculations have been performed for titanium dioxide in the anatase structure and
   zinc oxide  in the wurtzite structure.  We find that   the  quasi-stationary  electron distribution  function for ZnO  is  a  fermi-like curve that   rises linearly
   with increasing  excitation energy whereas the  analogous curve  for anatase   consists of a main peak  and a shoulder.  The calculations demonstrate  that
   the relaxation of  excited  electrons and  the setting  of the quasi-stationary distribution occur  within   the time  no more than 500 fsec for ZnO  and 100 fsec for anatase.
   We also discuss  the applicability  of  the effective  phonon model  with energy-independent   electron-phonon transition probability.   We find  that  the model  only
    reproduces  the trends  in  changing  of the characteristic times   whereas  the  precision of  such calculations is not high.  The  rate  of  energy  transfer  to phonons
   at  the quasi-stationary electron  distribution also have  been    evaluated  and  the effect  of  this transfer  on  the photocatalyses has been discussed.   We found  that
   for ZnO this rate  is about  5 times less than in anatase.
\end{abstract}

\pacs{}

 \maketitle

\section{Introduction}
   The  zinc oxide in  wurtzite  structure   and  the titanium dioxide  in  anatase structure are the  semiconductors of great
   interest  from  many viewpoints. Both demonstrate photocatalytic activity    in  the  UV  region of the sunlight spectra
   and are widely used   as   basic compounds   in attempts to create  the photocatalysts   active   in the visual  sunlight  that   can
   be used  for cleaning the environment  from
   organic pollutants and  pathogenic bacteria   \cite{Hoff95,Carp04,Lu08,Ni07,Gird09}.  They are  also  perspective  materials  for applications in solar  energetics  and
   random access memory  devices \cite{Mai04,Kim04,Lane04}.  The  photocatalytic  activity, magnetic properties   and
   charge  transport in these oxides to a great extent depend   on the  dynamics of the excited states,  therefore   in many   works the characteristics
   of   fast  electron  and hole dynamics   have  been studied.  In  the works   \cite{Yama99,Hend07,Wen07,Sun05}
   the relaxation  rates  of excited  electrons in  ZnO   via electron-phonon  coupling  have been estimated
   whereas    in the papers  \cite{Wata00,Wata01,Seki04,Hend04}  the  characteristics  of   relaxation  rates  for anatase  have been evaluated.

   The  comprehension  of  the results obtained in the experimental works  requires  an application of   some theoretical  models
   that  should incorporate   both  the   conditions of experiments and the  characteristics of  electron  dynamics
  determined  by the  properties   of the solids under study. Such models have  been successfully developed  and widely applied  for
  the excited electron  relaxation in metals   \cite{Knor02a,Zhuk02,Zhuk04}.
  In the case of semiconductors  this  aim is,  however,  far from being achieved, partly  because
  only  recently  the rigorous  methods for evaluations of  the  characteristics  of electron dynamics have been
  developed.   The main  phenomenon   responsible   for the relaxation  of  the low-energy  excited electrons in semi-conductors
  (whose   energy with respect  to the bottom of the conduction band is no more than the band gap) is  the  electron-phonon  coupling.
  A number of   first-principle method for the  calculations of the parameters of this  coupling  have been developed  that permit  to
  perform subsequent calculations for the  temporal  characteristics of  electron relaxation.    Possibly,  the most promising of  them
  is provided   by  the density-functional perturbation  theory  \cite{Zein84,Baro01}   implemented in the pseudo-potential Quantum
  Espresso (QE) computer code.  Recently  this approach  has been applied to the  evaluations  of  the characteristics of  electron relaxation  in
 ZnO  \cite{Zhuk10b},   anatase and rutile \cite{Zhuk10a}, germanium \cite{Tyut11},.   The calculations  have demonstrated
  a reasonable  correspondence  to experimental data
 and  helped  to  understand  some fine relations between the band structure  and electron  dynamics in the compounds.

 The shortage  of the evaluations  of the Refs.  \cite{Zhuk10b,Zhuk10a}  was  that  they had been  performed for a   case of a single  excited  electron in the
 conduction  band  that falls  down  to the bottom of the  band  loosing  the energy via the electron-phonon coupling.   More rigorous    estimations  of
 the electron dynamics  should   include also the  emergence  of new  electrons   excited  by  the external  light source  and also  the  effect  of   filling
 the energy level  under  consideration   with  the electrons falling  from the higher levels.  The  method of such evaluations  is proposed in the current paper.
 Although  the concrete  results are obtained  for ZnO and  TiO$_2$,   the  proposed  method does not have  features  preventing  its application  to different
 kinds of semi-conductors.   The method is based on the fully first-principle calculations  of the   electron-phonon coupling constants.   So it  does not  have
simplifying  assumptions on the energy  dependence of  the rate of electron-phonon relaxation.   It  is worth noting that the neglect  of  the energy  dependence
of  the electron-phonon   interaction  is a conventional  approximation,  as it is  e.g.  with the Fr\"{o}hlich's hamiltonian   \cite{DasS88,Tao93,Komi01,Jang07}.
So,  together  with the   characteristics  of  electron relaxation  in ZnO and TiO$_2$,   we  discuss also  the applicability  of the energy-independent  electron-phonon
coupling  and  find  that  in the current case  it is  of very  limited accuracy.

\section{Method of calculations}
  \subsection{Quasi-stationary electron distribution}
   We start with the following Boltzmann equation for the time-dependent distribution function
   $f(t,\epsilon)$ of excited  electrons (hereafter EDF) at the energy   $\epsilon$ above the bottom
   of the conduction band of a semiconductor (hereafter - excess energy)
   \begin{eqnarray}\label{eq1}
   \frac{\partial f(t,\epsilon)}{\partial t} = (1 - n(t,\epsilon)) \int_0^{\omega_m} n(t,\epsilon + \hbar\omega)
   F(\epsilon + \hbar\omega,\omega)d\omega -  \cr n(t,\epsilon)\int_0^{\omega_m}F(\epsilon,\omega)d\omega(1-n(t,\epsilon - \hbar\omega)) +
   \frac{\partial f_{inst}(t,\epsilon)}{\partial t}
   \end{eqnarray}
   In this equation $n(t,\epsilon) = f(t,\epsilon)/N(\epsilon)$, where  $N(\epsilon)$  is the density of electronic states, is the population of  a single band state  at the $\epsilon$ level.
   First integral describes, in a momentum-averaged  manner,  the process of coming of electrons
   to the level $\epsilon$  from all  the electronic states higher  by the energy $\hbar\omega$,
   the process accompanied by  the emission of phonons  whose   maximum  frequency is $\omega_m$.
   $F(\epsilon + \hbar\omega,\omega)$, the spectral function of  electron-phonon interaction,
   is the probability  for electrons to fall  from all the band states with energy
   $\epsilon+\hbar\omega$ to the states at the  energy $\epsilon$ emitting  phonons with energy $\hbar\omega$
   satisfying the conditions of  energy  and momentum conservation. Similarly, the
   second integral describes the process of electron transitions from the $\epsilon$ level  to the
   levels  lower by  the phonon energy $\hbar\omega$.  We  discuss only the case of a small  intensity  of irradiation,  so  we  assume  that  the
    $n(t,\epsilon)$ values  are small  and omit  the factors  $1-n(t,\epsilon)$.  The spectral function $F(\epsilon,\omega)$,
   satisfying  to energy and momentum conservation, has the form
   \begin{equation}\label{eq2}
   F(\epsilon, \omega)=\sum_{{\bf qkk'}}\sum_{nn'\sigma}\delta(\epsilon-e_{n{\bf k}})
   P(n{\bf k},n'{\bf k'},\sigma{\bf q})
   \delta_{{\bf k-k'\pm q}}\delta(\omega-\omega_{\sigma {\bf q}})
   \delta(\epsilon - \hbar\omega - e_{n'{\bf k'}})
   \end{equation}
   where $ P(n{\bf k},n'{\bf k'},\sigma{\bf q})$ is the probability  of a single electron
   transition  between the electronic band states $| n{\bf k}\rangle$  and
   $|n' {\bf k'} \rangle$
   accompanied by the emission or absorption  of  the photon with frequency
   $\omega_{\sigma {\bf q}}$.  The calculations  of  such spectral  function, which for the Fermi level in metals is proportional
   to the well-known Eliashberg function, were  extensively discussed in literature
   \cite{Maha90,Grim}.  In accordance to the  'golden Fermi rule', this probability,
    per unit of time, is determined as
   \begin{equation}\label{eq3}
     P(n{\bf k},n'{\bf k'},\sigma{\bf q})  = (2\pi/\hbar)
     |\langle {n\bf k}|\Delta V_{{\bf q}\sigma}  | n'{\bf k'}\rangle|^2
     \delta(e_{n{\bf k}}-e_{n'{\bf k'}})
   \end{equation}
  Here   the value
   $\Delta V_{{\bf q}\sigma}$  is the variation of the self-consistent
  potential in crystal caused by the displacement mode of the  phonon, and
  the value $\langle {n \bf k} |\Delta V_{{\bf q}\nu}  | {n'\bf k'} \rangle$
  is the matrix element of the electron-phonon
  interaction. The energy of phonon, small in comparison with the  energy of electronic states,
  is neglected here. The ways of calculating this matrix element also have
  been well discussed \cite{Baro01}.

  The  term  $\partial f_{inst}(t,\epsilon)/\partial t$ in   Eq. \ref{eq1} describes
  the instantaneous  electron distribution  in the conduction band produced by  the external
  source of excited electrons. One can  take for this  term the approximation
   \begin{equation}\label{eq4}
   \frac{\partial f_{inst}(t,\epsilon)}{\partial t} = S_0(t)S(\epsilon)
   \end{equation}
  where the factor $S_0(t)$, which in general can  be  time-dependent,  is  the concentration
  of  the excited electrons  determined by the  power of the light radiation,  and
  $S(\epsilon)$ is the instantaneous spectral function ( hereafter IEDF) which describes
  the  probability  for an excited  electron  to have  the excess  energy  $\epsilon$.
  Naturally,  IEDF should be normalized to unity: $\int_0^\infty S(\omega)d\omega = 1$.

  If one  linearizes   the energy  dependence  of the  $F(\epsilon + \hbar\omega,\omega)$  with
    $\omega$ as a small parameter   then Eq. \ref{eq1} is written in the form

   \begin{equation}\label{eq5}
   \frac{\partial f(t,\epsilon)}{\partial t} = \frac{\partial}{\partial \epsilon}
   [N^{-1}(\epsilon)f(t,\epsilon)B(\epsilon)] + \frac{\partial f_{inst}(t,\epsilon)}{\partial t}
   \end{equation}
   Here the function
  \begin{equation}\label{eq6}
    B(\epsilon) = \int_0^{\omega_{m}}\hbar\omega F(\epsilon,\omega)d\omega
  \end{equation}
  is the energy lost  by  the electrons at  the level $\epsilon$  via
  phonon emission. So the value  which we define as $\hbar \omega_0 = B(\epsilon)/N(\epsilon)$ is the averaged
  energy  lost  by $one$ electron in the process of  phonon emission;
  a first-principle  approach  to the calculations  of this value has been demonstrated in
  Refs. \cite{Zhuk10a,Zhuk10b}  (where it has been  denoted as $\Delta e$).

  Eq.  \ref{eq5} describes the time evolution  of  the $f(t,\epsilon)$ EDF when the  light
  source produces an instantaneous distribution of  excited electrons in the conduction band.
  We discuss here  the quasi-stationary case, when the temporal evolution  of  the external  pulse is slow and $dS_0(t)/dt << S_0(t)B(\epsilon)$. Taking $\partial f(t,\epsilon)/\partial t =0$
  one has the solution of  the Eq. \ref{eq5} in  the form
  \begin{equation}\label{eq7}
    f(t,\epsilon) = S_0(t)\frac{1}{\hbar \omega_0} \int_{\epsilon}^{\epsilon_{m}} S(\epsilon')d\epsilon'
  \end{equation}
  where $\epsilon_{m}$ is  the highest  excess energy  of  the excited electrons.

  The IEDF $S(\omega)$ can be obtained from the electronic band structure calculations. Namely,
  if the energy of the quantum  of optical excitation is $E_{exc}$, then  for the excess energy
  $\epsilon$ one has to sum the probabilities of all direct excitations from the electronic states at
  the energy $\epsilon -E_{exc}$ to  the  states at the energy $\epsilon$.
  Hence, the un-normalized IEDF is

   \begin{equation}\label{eq8}
   S(\epsilon)=\sum_{{\bf k}nn'}\delta(\epsilon-e_{n{\bf k}})
   T(n{\bf k},n'{\bf k})\delta(\epsilon - E_{exc} - e_{n'{\bf k}})
   \end{equation}
   where  $T(n{\bf k},n'{\bf k})$ is the probability  of the transition between the states
   $|n{\bf k}\rangle$ and $| n'{\bf k}\rangle$.
  (In practical calculations we replace the $\delta$-functions with the normalized to unity gaussians whose width  at  the half-maximum is 0.01 eV.)
  The transition probability  also can be evaluated basing on the first-order perturbation theory.
  In order to calculate  the matrix elements of  the Eq.  (\ref{eq8})  we take  advantage
  of  the atomic sphere  approximation \cite{Ande87}.     In  this approximation  the integration
  over  the space  of a crystal  is replaced with  the integration  over  atomic spheres.  In
  every  atomic sphere we take for  the perturbation  the dipole approximation  \cite{Loud83}. With this approximation the hamiltonian of
  the interaction of an atom  with  the electric field of light $\bf E(\omega)$ has  the form
   \begin{equation}\label{eq9}
     H_S = e{\bf D}_S{\bf E}(\omega)
   \end{equation}
   Here $e{\bf D}_S= e\sum_j{\bf r}_j^S$,  where  ${\bf r}_j^S$   is  the radius with respect
   to  the center of  the given atom,   is the operator  of  the dipole  moment  of  the atom.
   Hence,  the interaction  of an electron with  the field is
   \begin{equation}\label{eq10}
     H(\omega) = e  \sum_S {\bf r}^S{\bf \epsilon_E}E(\omega)
   \end{equation}
  We consider  the case of interaction of light with a polycrystal, so we have for the angle-averaged
  transition rate
    \begin{eqnarray}\label{eq11}
       T(n{\bf k},n'{\bf k}) = \cr
            \frac{2\pi}{\hbar } \cdot e^2 \cdot  (E(\omega))^2
              \delta( e_{n{\bf k}} -  e_{n'{\bf k}} -  \omega )
          \frac{1}{3} |\sum_S\langle {\bf k},i  |{\bf  r}^S | {\bf k},j   \rangle  |^2 \cr
   \end{eqnarray}
   Here the overlap  of  atomic  spheres is neglected,  and  the coefficient  1/3 emerges because  of  the averaging  over  the angle between  the vector ${\bf r}^S$  and  the directional  vector of  the field ${\bf \epsilon}_{\bf q}$,  see details in \cite{Loud83}.

  \subsection {Electron-phonon energy loss time  and the time of EDF  setting}
  The  value $F(\epsilon + \hbar\omega,\omega)$  in Eq. \ref{eq1}, according to the definition (\ref{eq2}),  is  the probability
   of  transitions from $all$  the electronic states at  the energy $\epsilon + \hbar\omega$ to $all$ the states at the energy $\epsilon$. So  we apply
  to this value  the approximation
   \begin{eqnarray}\label{eq12}
    F(\epsilon + \hbar\omega,\omega) =  N(\epsilon + \hbar\omega)P(\epsilon)
                                        N(\epsilon)\delta(\omega - \omega_0)
   \end{eqnarray}
   where  $P(\epsilon)$  is  the momentum-averaged probability  of
   a $single$  transition  at the electron energy $\epsilon$ and  phonon energy $\hbar\omega_0$.   The energy dependence of  $P$  is often neglected,
   as it takes place in  numerous  works with  Fr\"{o}hlich's  electron-phonon interaction, see e.g.
   \cite{DasS88,Komi01,Jang07}. We  will  show  in the next  section  that
   the energy dependence of $P(\epsilon)$ can  be  omitted only  for the aims of interpretations.  In fact this dependence is not negligible.  Employing approximation (\ref{eq12})   and
   introducing  for the external  source  definition
   \begin{eqnarray}\label{eq13}
    \frac{\partial n_{inst}(t,\epsilon)}{\partial t} = \frac{\partial f_{inst}(t,\epsilon)}{\partial t}/N(\epsilon)
   \end{eqnarray}
 we can rewrite  Eq. (\ref{eq1})  in the form
   \begin{eqnarray}\label{eq14}
    \frac{\partial n(t,\epsilon)}{\partial t} =
    n(t,\epsilon+\hbar\omega_0)P(\epsilon)N(\epsilon+\hbar\omega_0)
    - n(t,\epsilon)P(\epsilon)N(\epsilon-\hbar\omega_0)] +
    \frac{\partial n_{inst}(t,\epsilon)}{\partial t}
   \end{eqnarray}
   After   linearization near the energy  $\epsilon$ the equation becomes
   \begin{eqnarray}\label{eq15}
    \frac{\partial n(t,\epsilon)}{\partial t} =
    P(\epsilon) \hbar \omega_0 \frac{1}{N(\epsilon)}
    \frac{\partial}{\partial \epsilon}[N^2(\epsilon)n(t,\epsilon)]
    +
    \frac{\partial n_{inst}(t,\epsilon)}{\partial t}
   \end{eqnarray}
   We  introduce  a  new variable
   \begin{eqnarray}\label{eq16}
     q(t,\epsilon) = N^2(\epsilon)n(t,\epsilon)
   \end{eqnarray}
   and  get for this  value   the equation
   \begin{eqnarray}\label{eq17}
    \frac{\partial q(t,\epsilon)}{\partial t} =
    \hbar\omega_0 P(\epsilon) N(\epsilon)\frac{\partial q(t,\epsilon)}{\partial \epsilon} +
    \frac{\partial n_{inst}(t,\epsilon)}{\partial t}
   \end{eqnarray}
One can show  that in the absence of  light the formal solution of Eq.  (\ref{eq17})  at the energy $E$ can  be written    as
   \begin{eqnarray}\label{eq18}
    q(t,E) = \Phi \{ t + [\hbar\omega_0]^{-1} \int_{E_i}^E  d\epsilon (P(\epsilon)N(\epsilon))^{-1}
   \end{eqnarray}
  where  $\Phi$  is an arbitrary  function;  this function and  the  energy  $E_i$ have  to  be chosen  in order to satisfy  initial  conditions.
  We  assume for the  initial conditions  that at $t=0$ the electron  is excited by the external  source  to the level $E_i$, and  afterwards  the source is switched off.
   It is easy  to check that  these conditions are satisfied if     $\Phi$  is  a $\delta$-like function, so
   \begin{eqnarray}\label{eq19}
    n(t,E) =  N^{-2}\delta \{ t + [\hbar\omega_0]^{-1} \int_{E_i}^E  d\epsilon (P(\epsilon)N(\epsilon))^{-1}\}
   \end{eqnarray}
  The equation
   \begin{eqnarray}\label{eq20}
     t + [\hbar\omega_0]^{-1} \int_{E_i}^E  d\epsilon (P(\epsilon)N(\epsilon))^{-1} = 0
   \end{eqnarray}
   is then  that of  relaxation  of the excited electron.  Hence  for  the  rate  of  the  energy  relaxation  we have
   \begin{eqnarray}\label{eq21}
   dE/dt = - \hbar\omega_0P(E) N(E)
   \end{eqnarray}
   and the energy-loss time,  that is the time necessary  for the electron  to fall from the level  $E_i$ to the bottom  of  the conduction band, is
   \begin{eqnarray}\label{eq22}
   \tau_l(E_i) =  (\hbar\omega_0)^{-1} \int_0^{E_i}(P(\epsilon) N(\epsilon))^{-1} d\epsilon
   \end{eqnarray}
  In real  calculations the lower limit of integration  has to be  replaced with the  maximum energy of phonons $\hbar \omega_m$ since  at  lower
  energy  the distribution function  is determined  by  different  mechanisms, mainly by  the  electron-hole  recombination.

  Eq.  (\ref{eq22})  demonstrates the relation  between the   time of  energy  loss and the density of states. One can  reveal  also  the  relation
  between the energy-loss  time in the current paper and  that   in the previous works \cite{Zhuk10a,Zhuk10b}.    In   the  cited papers  the rate
  $\Gamma(\epsilon)$  of the  electron-phonon relaxation   has been defined  via
   \begin{equation}\label{eq23}
   \Gamma(\epsilon) = P(\epsilon)N(\epsilon).
   \end{equation}

    So  the  $\Gamma$ value  is the transition  probability from  the  electronic state  at energy $\epsilon$ to $all$  the states inside   the  energy   interval  from   $\epsilon -\hbar\omega_m$   to  $\epsilon$, and  the equation  for the energy-loss time is
   \begin{eqnarray}\label{eq24}
   \tau_l(E_i) =  (\hbar\omega_0)^{-1} \int_0^{E_i} \Gamma^{-1}(\epsilon) d\epsilon
   \end{eqnarray}
   Since the value  $\hbar \omega_0  \equiv \Delta e$  is almost  energy-independent \cite{Zhuk10a,Zhuk10b},   this equation   is equivalent
     to  the  equation  for the energy-loss time
   \begin{eqnarray}\label{eq25}
   \tau_l(E_i) =   \int_0^{E_i} (\Delta e (\epsilon)\Gamma(\epsilon))^{-1} d\epsilon
   \end{eqnarray}
   proposed  in the papers \cite{Zhuk10a,Zhuk10b}.

    One more function of interest  is  also   the time of setting of the quasi-stationary electron distribution.
    Now  we assume   that the   source of light   is  switched on
     at t=0  and   slowly  changes afterwards   satisfying   the condition  $dS_0(t)/dt << S_0(t)B(\epsilon)$.    The time of setting
     is determined  as   the time necessary  for   transient processes  to extinct  after  the light is switched on.
    Employing definitions (\ref{eq13},  \ref{eq23})   and neglecting the energy  dependence of $P$ we can rewrite Eq. (\ref{eq14}) as
   \begin{eqnarray}\label{eq26}
    \frac{\partial n(t,\epsilon)}{\partial t} =
    n(t,\epsilon+\hbar\omega_0)\Gamma(\epsilon+\hbar\omega_0)
    - n(t,\epsilon)\Gamma(\epsilon-\hbar\omega_0) +
    \frac{\partial n_{inst}(t,\epsilon)}{\partial t}
   \end{eqnarray}
    We  solve  this equation supporting   on a set  of  excess electron energies  $\epsilon_p =\hbar\omega_m + p\hbar\omega_0$, p = 0 $\div$  m,
    that is from $\hbar\omega_m$  to  the maximum excess energy $\epsilon_m$.  Introducing  the notation
    $\Gamma_p = P(\epsilon_p) N(\epsilon_p)$  we transform  Eq.  (\ref{eq28})  to  the set of  equations near the $\epsilon_p$  levels:
    \begin{eqnarray}\label{eq27}
    \frac{\partial n(t,\epsilon_m)}{\partial t} =
    - n(t,\epsilon_m)\Gamma_{m-1} +
    \frac{\partial n_{inst}(t,\epsilon_m)}{\partial t}
    \cr
    \frac{\partial n(t,\epsilon_{m-1})}{\partial t} =
    n(t,\epsilon_m)\Gamma_{m}
    - n(t,\epsilon_{m-1})\Gamma_{m-2} +
    \frac{\partial n_{inst}(t,\epsilon_{m-1})}{\partial t}
    \cr
    ... \hspace{50mm}
    \cr
   \frac{\partial n(t,\epsilon_{m-p})}{\partial t} =
    n(t,\epsilon_{m-p+1})\Gamma_{m-p+1}
    - n(t,\epsilon_{m-p})\Gamma_{m-p-1} +
    \frac{\partial n_{inst}(t,\epsilon_{m-p})}{\partial t}
    \cr
    ... \hspace{50mm}
    \cr
   \frac{\partial n(t,\epsilon_0)}{\partial t} =
    n(t,\epsilon_1)\Gamma_1
   +     \frac{\partial n_{inst}(t,\epsilon_0)}{\partial t}
   \end{eqnarray}
   For the highest  energy  level the  solution   is
   \begin{eqnarray}\label{eq28}
     n(t,\epsilon_m) =
   [ \int_0^t  \frac{\partial n_{inst}(t',\epsilon_m)}{\partial t}e^{\Gamma_{m-1}t'}  + n(0,\epsilon_m)]e^{-\Gamma_{m-1}t}
   \end{eqnarray}
   For the  lower  energy  levels, except of  the 0-th,   the solution obtained by  recursion  is
   \begin{eqnarray}\label{eq29}
     n(t,\epsilon_{m-p}) =
   [ \int_0^t  \frac{\partial \tilde n_{inst}(t',\epsilon_{m-p})}{\partial t}e^{\Gamma_{m-p-1}t'}  + n(0,\epsilon_{m-p})]e^{-\Gamma_{m-p}t}.
   \end{eqnarray}
    Here  we defined  the  modified  source  function
   \begin{eqnarray}\label{eq30}
    \frac{\partial \tilde n(t,\epsilon_{m-p})}{\partial t } =
   \frac{\partial{n_{inst}(t,\epsilon_{m-p})}}{\partial t}  + n(t,\epsilon_{m-p+1})\Gamma_{m-p+1}
   \end{eqnarray}
   The solution  for the m-th level  has   terms proportional  to $e^{-\Gamma_{m-1}t}$,  so  omitting    these terms decaying with
   characteristic time  $\tau_m = 1/\Gamma_{m-1}^{-1}$  one obtains static solution
 \begin{equation}\label{eq31}
    n(t,\epsilon_m)  =  \frac{\partial{n_{inst}(t,\epsilon_m)}}{\partial t}/\Gamma_{m-1}
 \end{equation}
    The solution  for the level  $\epsilon_{m-1}$   becomes  static after the time
 \begin{equation}\label{eq32}
    \tau_{m-1} = \tau_m + 1/ \Gamma_{m-2}
 \end{equation}
   and it has  the form
 \begin{equation}\label{eq33}
    n(t,\epsilon_{m-1})  =  \frac{1}{\Gamma_{m-1}\Gamma_{m-2}}
   [ \frac{ \partial n_{inst}(t,\epsilon_m)} {\partial t}  + \frac{\partial n_{inst}(t,\epsilon_{m-1})}{\partial t}\Gamma_{m-1}]
 \end{equation}
 It  follows from further recursion that for an  arbitrary level  $\epsilon_p$  the static  solution  is realized after  the time
 \begin{equation}\label{eq34}
    \tau_p = \sum_{s=p}^{m-1} 1/ \Gamma_s
 \end{equation}
and it is
 \begin{equation}\label{eq35}
    n(t,\epsilon_p)  =  \frac{1}{\Gamma_p\Gamma_{p-1}}
  \sum_{s=p}^{m-1} \frac{ \partial n_{inst}(t,\epsilon_s)} {\partial t} \Gamma_s
 \end{equation}
    Taking into account the smallness of  the $\hbar\omega_0$  value  and  the definition (\ref{eq23})  we can
    replace  these  sums with  integrals
 \begin{equation}\label{eq36}
    \tau_s (E_i) =  (\hbar\omega_0)^{-1}\int_{E_i}^{\epsilon_m}\Gamma(\epsilon )^{-1}d\epsilon
 \end{equation}
and
 \begin{equation}\label{eq37}
    n(t,E_i)  =  \frac{1}{\hbar\omega_0(P(E_i)N(E_i))^2}\int_{E_i}^{\epsilon_m}P(\epsilon)N(\epsilon )\frac{ \partial n_{inst}(t,\epsilon )} {\partial t}d\epsilon
 \end{equation}
 Eq. (\ref{eq36}) is   similar  to  Eq. (\ref{eq22}), but  contrary  to  the time  of  energy loss    the time of EDF setting  is
 determined by the time of  the transitions  to $E_i$ from all  the $higher$   states.  It is easy  to show  that if the source  function is  defined  via   Eq.
 (\ref{eq4})  the Eq.  (\ref{eq37})   becomes, as it  has to be,  equivalent  to   Eq. (\ref{eq7}).

\section{Technical details}
   We apply  the described approach to the cases of ZnO in the structure of wurtzite and TiO$_2$  in the anatase structure.
   Pure ZnO  is  the semiconductor  with  direct band gap of  3.4 eV, whereas anatase has the band gap 3.2 eV wide.
   The energy band  structure  of the pure  anatase and zinc oxide has been  extensively studied, mainly  by the methods of  the density
    functional theory. Normally such methods produce  the value of band
   gap  much less  than  the experimental data.  The present  study is based on the band
   structure calculations \cite{Zain10a,Zhuk10a,Zhuk10b} for the pure  and doped anatase  modified  by  applying  single-site coulomb
   correlation corrections within the LSDA+U approach based on the LMTO band structure method.
   Such approach  produces  for the pure and doped anatase  the values  of the  band gap and  the energies of impurity states
     close  to experimental data.  However, this method of the band gap correction  is not  sufficient  for zinc oxide.
     So in this case we  apply to the conduction band states of  zinc oxide the 'scissor operator', hence  we
      perform a rigid  shift of all the conduction band states to  higher energy  until  a good  value of  the band
       gap is achieved. Such approach  is justified by  the comparison  with  the results of  the band structure
       calculations of ZnO corrected  by  the application of  the GW many-body  theory \cite{Usud02}.  They demonstrate that the
     application   of  the many-body GW corrections  produces  almost uniform shifts  of  the conduction  band states to higher energy.

\section{Results and discussions}
\begin{figure}[]
\includegraphics [width=220pt, height=200pt]{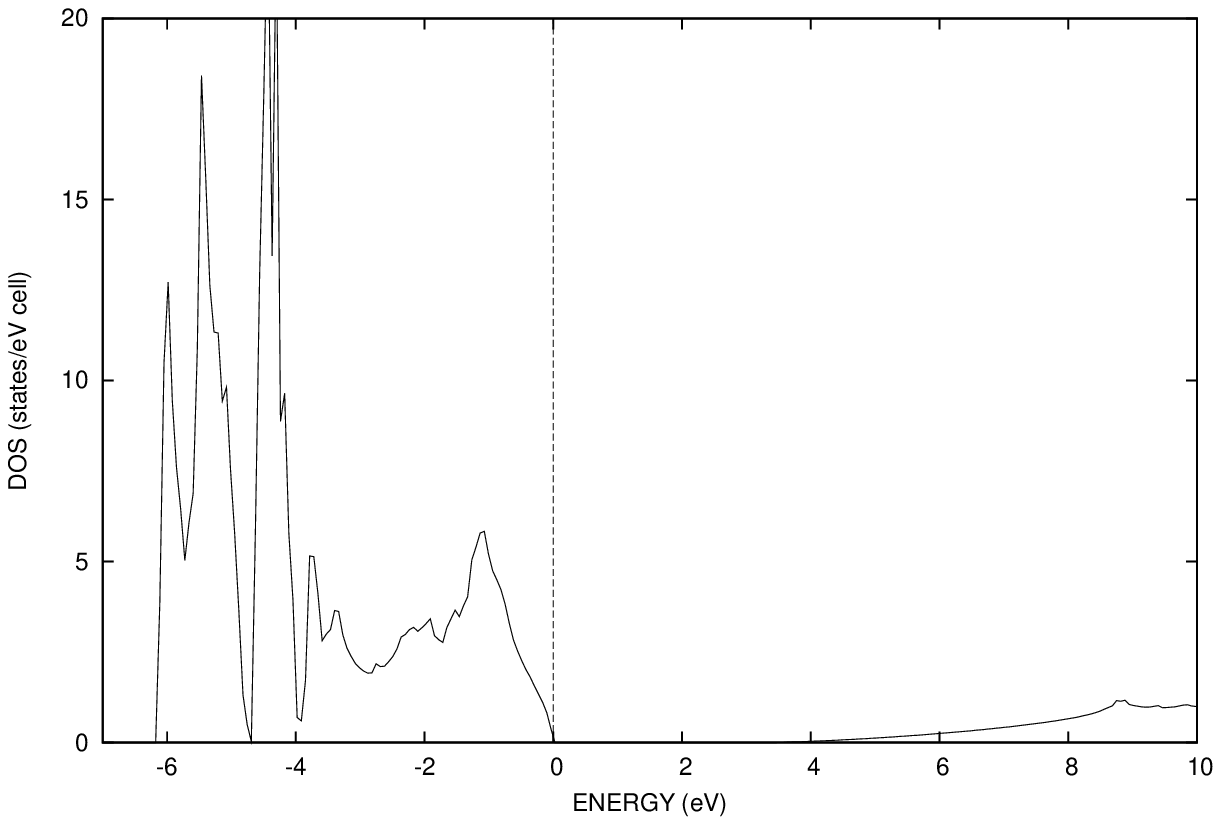}
\includegraphics [width=220pt, height=200pt]{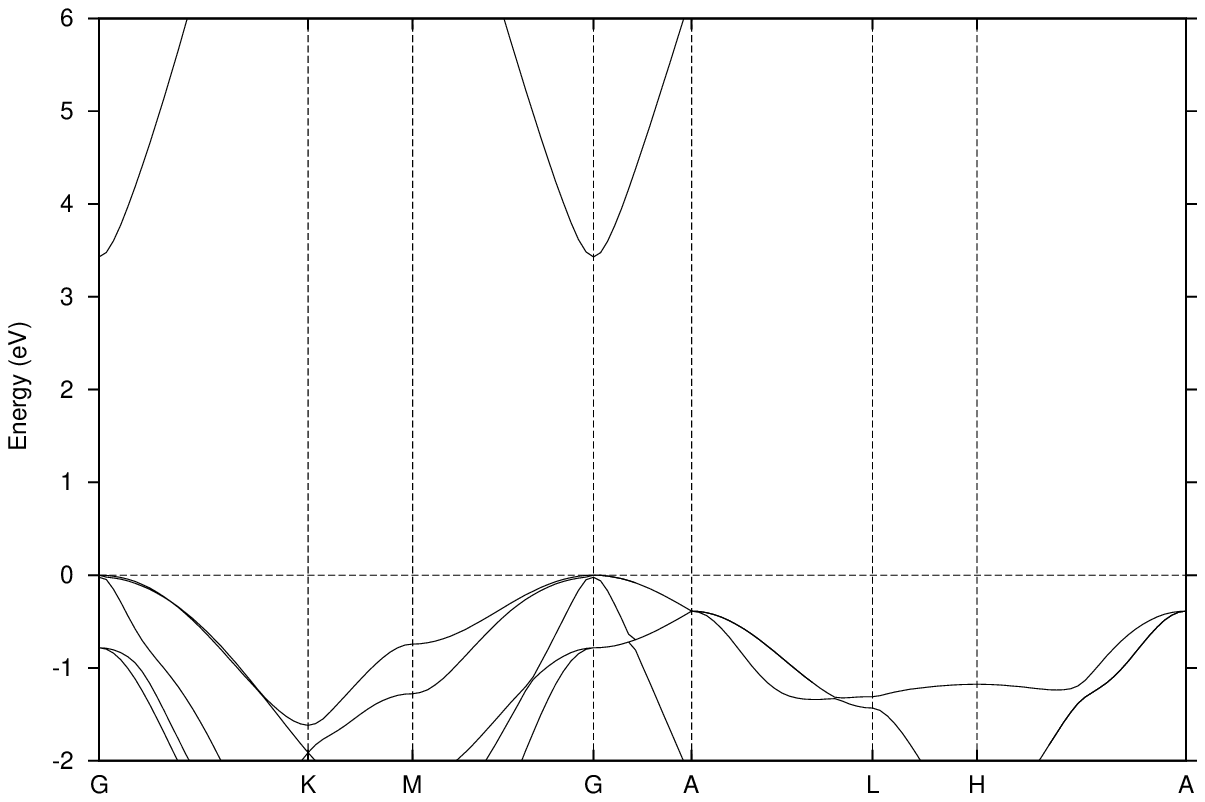}
\caption {Left panel: the density of states for ZnO; right panel: electron dispersion curves  of  zinc oxide near the band gap.  Both kinds of data are  given  with
 respect  to the Fermi level.
          }
\label{Fig1}
\end{figure}

\begin{figure}[]
\includegraphics [width=220pt,height=200pt]{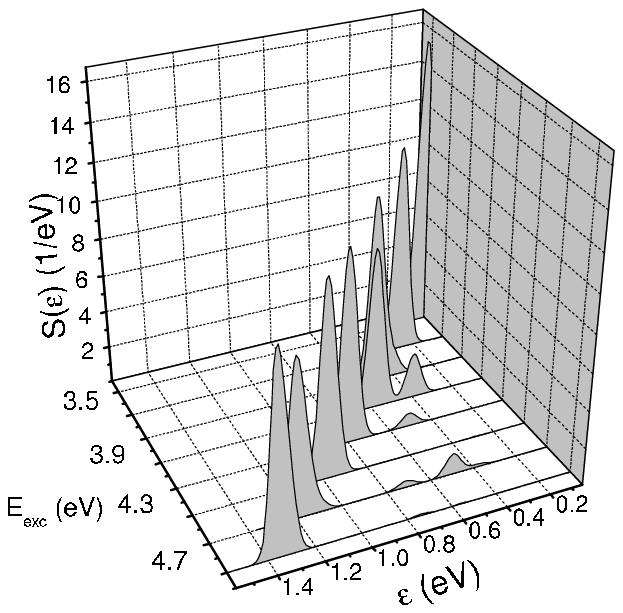}
\includegraphics [width=200pt,height=200pt]{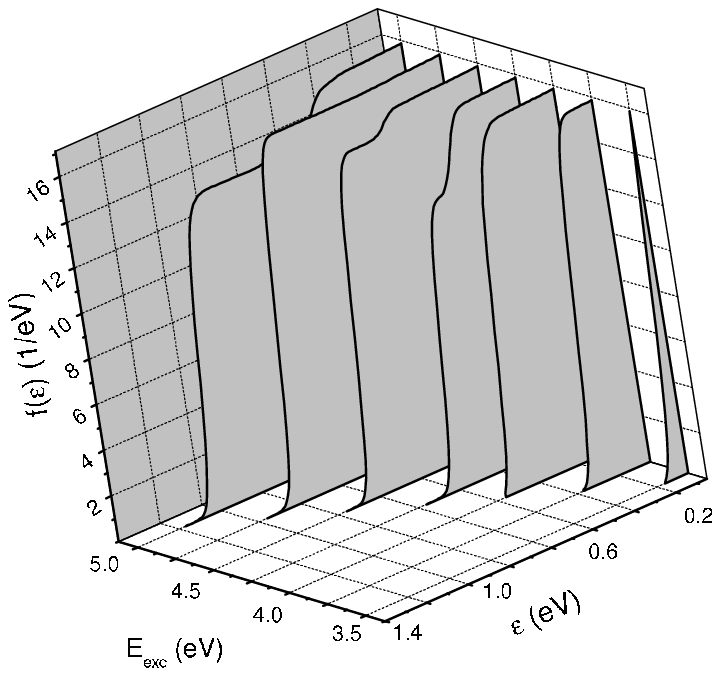}
  \caption { IEDF $S(\epsilon)$ and EDF $f(\epsilon)$ values for  zinc oxide given as  functions
           of  the  excess energy of excited electrons $\epsilon $  and
          excitation energy   $E_{exc}$ .
          }
  \label{Fig2}
  \end{figure}

  In  Figs. \ref{Fig1} and \ref{Fig2}  the main  data on the band  states,  IEDF  and EDF  are given  for zinc oxide.    The  bands of ZnO between  -6.2  and  - 4 eV
   are those composed  of the  Zn 3d-states, the states  between   -4 eV and the Fermi level consist of the  O 2p  valence  states,
    whereas   the conduction  band states, composed  of the Zn  4s-states,  are separated  by the direct  band gap    3.4 eV  wide.
 \begin{figure}[]
\includegraphics [width=200pt,height=180pt]{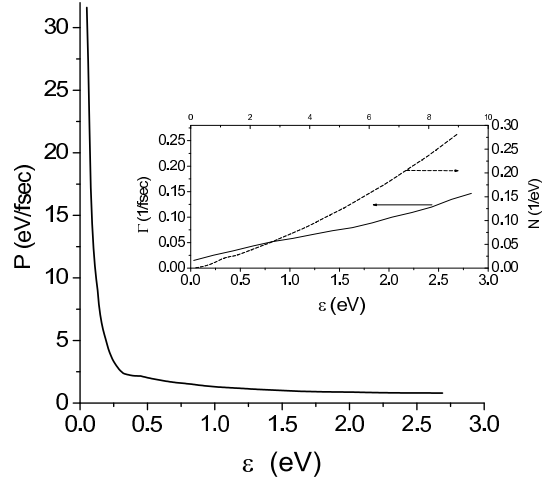}
  \caption {Dependence  of  the  relaxation rate  P   on the excess energy $\epsilon$ for zinc oxide.  In the inset  also  the energy dependencies of  the
               total relaxation rate  $\Gamma$  and density  of states  N are given.
          }
  \label{Fig3}
  \end{figure}
 \begin{figure}[]
\includegraphics [width=200pt,height=180pt]{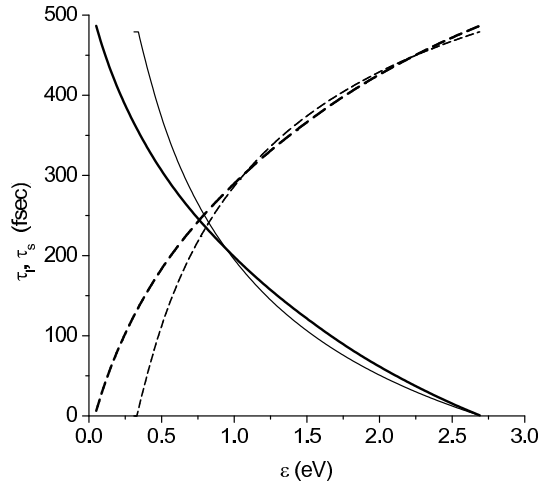}
  \caption {Dependence  of  the  EDF setting  time $\tau_s$  (thick solid line) and energy loss time $\tau_l$ (thick dashed line) on the excess energy
  $\epsilon$ for ZnO.  Also  the values of the $\tau_l$   (thin solid line) and $\tau_s$ (thin dashed line) are given calculated  with  the constant
  transition probability P= 1 eV/fsec.
          }
  \label{Fig4}
  \end{figure}

  It follows from  the Fig. \ref{Fig2}  that in the case of zinc oxide the main  feature of the  $S(\omega)$  IEDF  function is  a gaussian peak whose  maximum energy
   increases  linearly  with the rise of  the excitation energy. This peak corresponds  to the vertical  transitions
  from  the highest valence band states to the lowest  conduction band states   for the points located  at  the  $\Gamma-K$
  direction  of  the Brillouine zone  and  in the vicinity of this direction.  With  the increase  of  the excitation energy the wave
  vector of  the  transition  shifts from the $\Gamma$ to K point. The  transition matrix elements of  such excitations are  much higher  than  the TME  for  excitations
  from all the lower  valence band states,  so  the excitations from the lower  states manifest  themselves as  the low satellites of  the main peak.
  In correspondence  with  this IEDF  function,  the quasi-stationary  function $f(\omega)$  demonstrates a Fermi-like distribution  of  the excited  electrons in the conduction
  band, with almost constant number of electrons at the energy levels   from the bottom of the conduction band to the energy  $E_{exc}-3.4$ eV.

  In   Figs. \ref{Fig3} and \ref{Fig4}  the data concerning electron dynamics in ZnO anatase are given.   Fig. \ref{Fig3}  also characterizes the applicability  of  the
   "effective phonon"  approximation with energy-independent transition probability $P$.  The   change of the total probability
  $\Gamma(\epsilon)$  with energy  is almost   linear  beginning  from the lowest  excess energy $\epsilon = \hbar\omega_m$, whereas   the density  of states $N(\epsilon)$  also
  changes  almost  linearly,  but  reduces to very small values  at  the low  limit.  Therefore the  value $P(\epsilon)=\Gamma(\epsilon)/N(\epsilon)$  diverges  near
  the bottom of  the     conduction band, but  at  the energy  above  0.3 eV the   change  of  $P(\epsilon)$   with energy is slow.

    The calculated data on the energy loss time  $\tau_l$ and  EDF   setting time $\tau_s$  are shown  in Fig. \ref{Fig4}.  They  demonstrate  that   the relaxation
    of the  electrons   to the bottom of  the conduction band   occurs within the  time less than  500 fsec.  The  setting  of  the
    quasi-stationary electron distribution   in presence of the steady in time light source  also occurs within the time of no more than 500 fsec.  These  data are in reasonable
    correspondence    to  the experimental data \cite{Yama99,Sun05,Hend07,Wen07}.  An analysis  of the problems of  comparing  the experimental  and  theoretical
    data  can be found in  the Ref. \cite{Zhuk10b}.

    In Fig.  \ref{Fig4} also the  data on  $\tau_l$, $\tau_s$    are given calculated  with a constant  value of the transition probability  $P$. In  order to obtain
    the best  results the  integration interval  from  0  to 0.3  eV has been  from the calculations excluded.    The calculations demonstrate   that  the approximation of  the
    energy-independent  P  produces  sufficiently  good  results  only at  the excess energy more than 0.7 eV.  Variation  of  the  $P$-value does not help to improve  the
    correspondence  to the results of  the calculations with  the energy-dependent probability.

\begin{figure}[t]
\includegraphics [width=220pt,height=200pt]{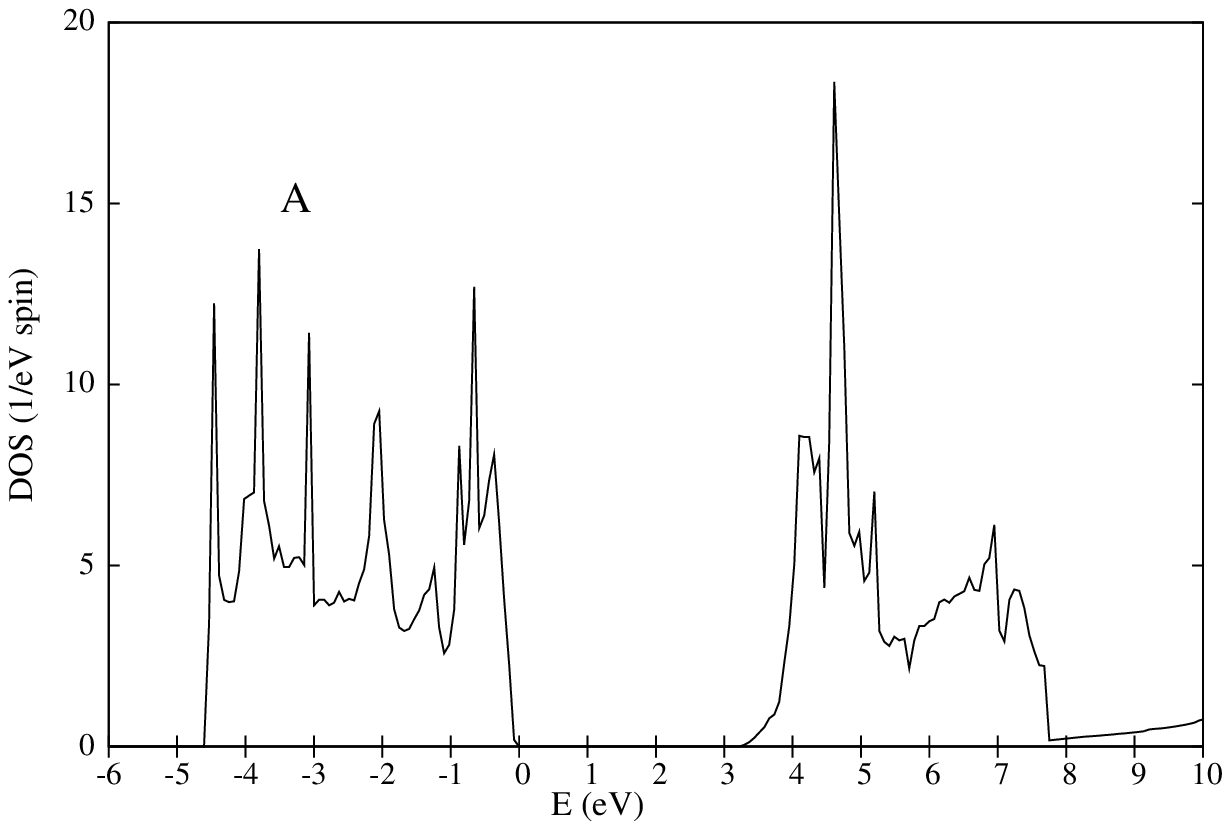}
\includegraphics [width=220pt,height=200pt]{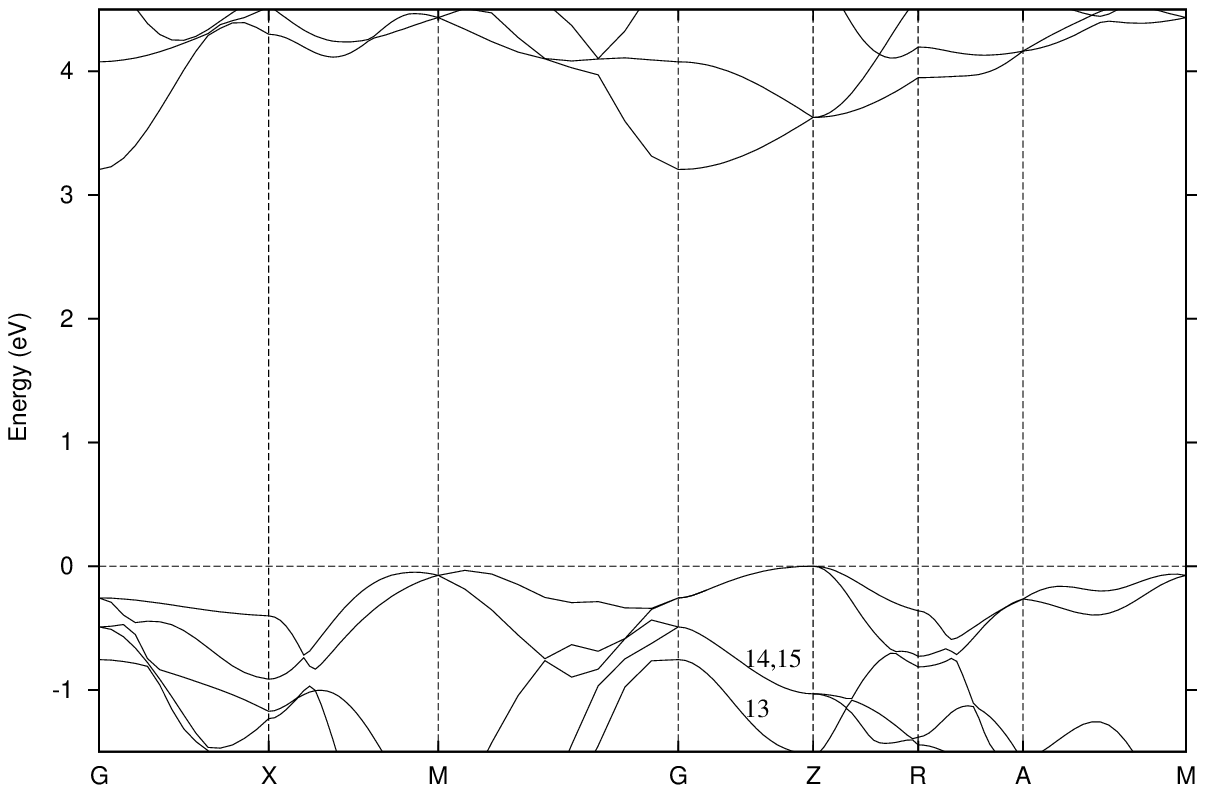}
  \caption {Left panel: the  density   of  states of anatase;  right panel: energy band structure of anatase  near the band gap.
  Both kinds of data are  given  with respect  to the Fermi level.
          }
  \label{Fig5}
  \end{figure}

 In Fig. \ref{Fig5}  the  density of states and dispersion curves for anatase  are given. The  conduction  band states  at  the energy  from 3.2 to 7.8 eV above
 the Fermi level are composed mainly of 3d Ti states.  The corresponding   density  of states sharply
  changes with energy  that evokes essential  variation of  $\Gamma$-probability
 \cite{Zhuk10b}.  On opposite  to the case of ZnO  the calculations demonstrate  that  the band  gap of anatase  is  not direct.   The highest  valence band state with almost
 equal energy are observed  in $Z$- and  $M$-point  whereas the lowest  conduction band  state is in $\Gamma$-point.  The decrease of energy
 of the  highest  valence  states   on direction from $M$ to $\Gamma$  points  has been  confirmed  recently  on ARPES experiments \cite{Emor10}.
\begin{figure}[]
\includegraphics [width=200pt,height=200pt]{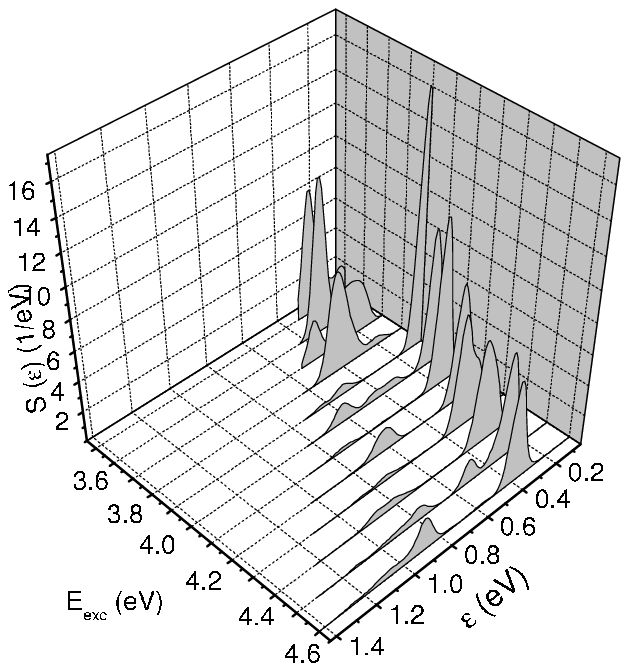}
\includegraphics [width=200pt,height=200pt]{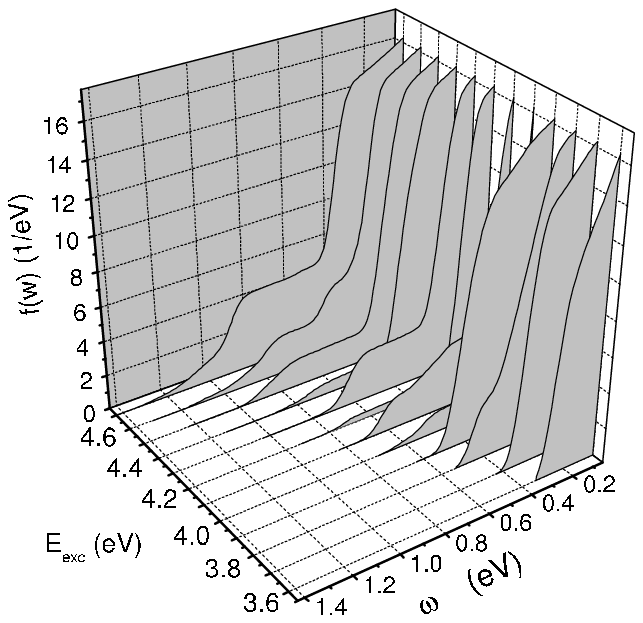}
  \caption {The dependencies  of the IEDF $S(\epsilon)$ and EDF $f(\epsilon)$  functions
           on  the  excess energy of excited electrons $\epsilon $  and
          excitation energy   $E_{exc}$ for  anatase.
          }
  \label{Fig6}
  \end{figure}

  At the excitation energy from  3.5  to 3.8  eV   the $S(\omega)$ dependencies for anatase also are gaussian peaks
  accompanied  by  low-energy satellite. Similarly  to the the case of zinc oxide  they correspond to  the transitions from  the highest  occupied
  valence band states  to the lowest conduction band states.  The  wave  vectors of these  excitations are located  on
  the $\Gamma-Z$ direction or  in the vicinity  of  it.  For  example, the excitations at 3.5 eV  occur from the states near the middle point  of  the
  $\Gamma - Z$  direction,  the excitations  at  3.6 eV occur from the states near $Z$ point,
  and  the  excitations at 3.8 eV  occur from the states near the $\Gamma$ point.
  However,  contrary to  the  case of zinc oxide, the TME  for excitations from  some lower band states are higher than  those
   for the  excitations from  the highest  valence band states.  Such  are  the band states that belong  to the  13-th and
   14,15-th degenerate  dispersion curves on Fig. \ref{Fig4}  and  the states of general  symmetry  in the vicinity of  these curves.
  So,  when  the excitation  energy increases up to  the value   sufficient  for the excitations from these lower bands, and this energy is of 3.9 eV, a shift  of
  the $S(\omega)$ peaks to the lower $\omega$ occurs.   At   energy above this threshold  the excitations  from the highest
  valence band  states  to  the lowest  conduction band states  contribute to only low satellites  of  the the $S(\omega)$  functions.  In comparison with ZnO,
  this modification   of  $S(\omega)$  provokes
  essential change  in the corresponding $f(\omega)$ EDF function.  At  energy  above 3.9 eV  it  is no more a Fermi-like  regularity,  but is composed of the main peak  at
  $\epsilon$ $\leq$0.6  eV  and a  low shoulder  extending up   to the excess  energy equal to  $E_{exc}-3.2$ eV. So  at  any $E_{exc}$   and $\epsilon$  the number  of excited electrons in
  anatase is markedly less than  in ZnO.
\begin{figure}[]
\includegraphics [width=220pt,height=200pt]{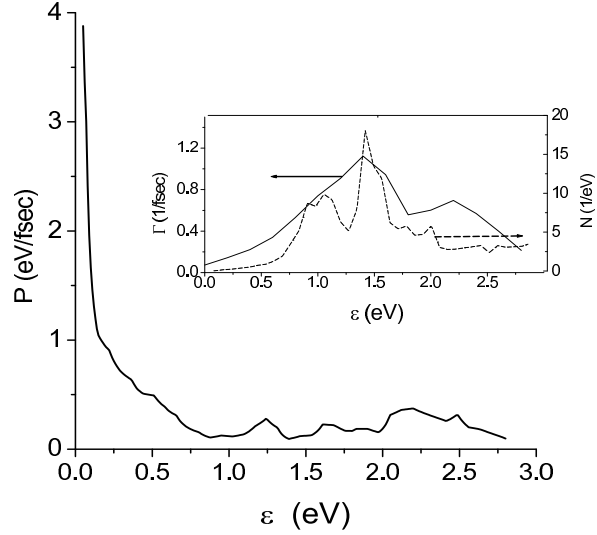}
  \caption {The  dependence  of  the  relaxation rate  P   on the excess energy $\epsilon$ for anatase.  In the inset   also  the energy dependencies of  the
               total relaxation rate  $\Gamma$  and density  of states are given.
                }
  \label{Fig7}
  \end{figure}
\begin{figure}[]
\includegraphics [width=200pt,height=180pt]{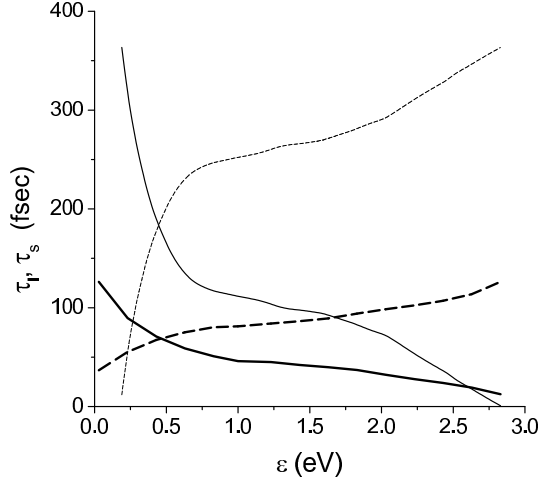}
  \caption {The  dependence  of  the  EDF setting  time $\tau_s$  (thick solid line) and energy loss time $\tau_l$ (thick dashed line) on the excess energy
  $\epsilon$ for anatase.  Also  the values of the $\tau_l$   (thin solid line) and $\tau_s$ (thin dashed line) are given calculated  with  the constant
  transition probability P= 0.25 eV/fsec.
          }
  \label{Fig8}
  \end{figure}

    The  characteristics of  electron dynamics are given for anatase in Figs.  \ref{Fig7}  and \ref{Fig8}.  As  in the case  of ZnO,  the transition probability   $P$ has  divergence
    near  the bottom  of  the conduction band. The  interval of  rapid change of  $P$  with  energy  extends from   zero  to about  0.7 eV.  At  higher  energy
    $P$   varies near  0.25  eV/fsec, the value much  less than  in the case of  ZnO.    Nevertheless, as it  follows from  Fig. \ref{Fig8},  both  the energy-loss time $\tau_l$
    and EDF setting time $\tau_s$ appear  to be much less than  for ZnO.  This  is associated   with much higher  values  of  density of state  that  provide
    higher  rate  of  electron-phonon relaxation.    Also  the $\tau_l$ and $\tau_s$  values  are  shown in Fig. \ref{Fig8}  calculated  with constant value  $P$=0.25 eV
    and with  interval  of rapid variations of  $P$  excluded from  integration.  It  is evident  that these  calculations only  reproduce  the trends of  changes,   whereas
    the  actual deviations  from the  exact  $\tau_l,\tau_s$-values mounts to 400 \%.  The variations of  the P-value   between 0.1 and 0.4 eV/fsec  and also of the low limit
    of integration do not help to  improve  these results.

\section{Conclusions}

   We  have proposed a first-principle method for  evaluations of   the   distribution  function of the excited  electrons in the  conduction band of
    semi-conductors.  The approach takes  into  account the loss of electron energy via electron-phonon  coupling,  the emergence  of  new electrons excited by
    the external  light source  and redistribution of electrons between the energy levels.  The  method  has been applied to  the evaluations  of the static  electron
    distribution function in anatase and zinc oxide;  the time of energy loss and the time of setting  of the static distribution also  have been calculated.

\begin{figure}[]
\includegraphics [width=300pt,height=180pt]{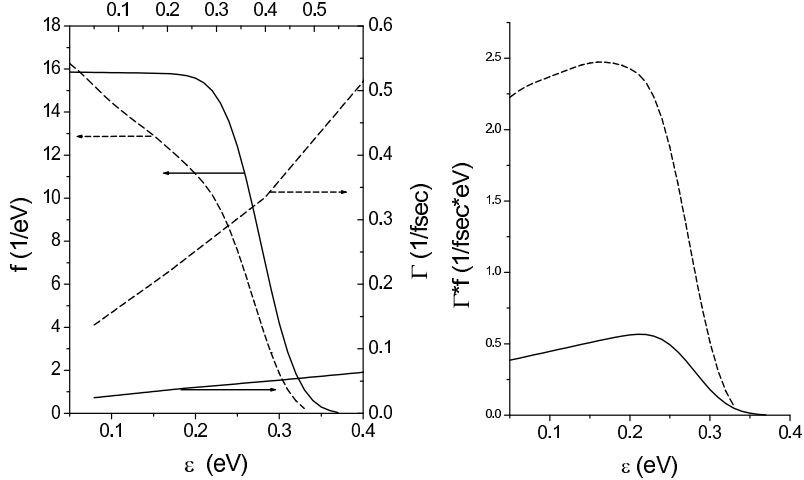}
  \caption {Left  panel: the  energy dependencies of  the $f(\epsilon)$  and $\Gamma(\epsilon)$ values  for  ZnO and anatase;
                  right panel: the energy dependencies of  the $\Gamma(\epsilon)f(\epsilon)$  products.  The data on ZnO are given in
                  solid curves,  and the data for anatase are in dash.
          }
  \label{Fig9}
  \end{figure}

    The  method helps   to come also to some conclusions  that  may have relation  to the photocatalytic  activity  of the compounds.
      It  is generally accepted  that  the photocatalytic activity  of the   oxide  semiconductors  is to a great  extent determined  by  the light   absorption
  because it  is proportional  to the number of created electron-hole pairs.
  Therefore the band  structure  and   optical  absorption in  pure and doped ZnO  and  TiO$_2$  was  a subject
  of a great number of calculations, see e.g.  Ref.   [\onlinecite{Cui08}].   Also a factor  important  for the photocatalytic  activity  is the
  rate  of  charge  transfer  between the oxide and molecules absorbed  on surface; this rate    should be higher than  the rate  of electron-hole recombination.
  This    factor  deserves a special attention,  but  till  now it was  only  hardly  touched  in  the first-principle approaches.  Our calculations permit to discuss  one more
  characteristic of oxides  that also can be important.  Namely,  one  should expect that,  irrespectively  of the details of the processes on the surface,   the  less  is the energy  loss of excited  electrons in bulk
  the more is the portion of  the absorbed  energy of light  that can be spent  to produce a photochemical  reaction.    So  it is worthwhile to introduce   the  value
 \begin{equation}\label{eq38}
     e_{ph}(\epsilon) = \hbar\omega_0 f(\epsilon)\Gamma(\epsilon)
 \end{equation}
   as the   measure of the energy  transfer  to the phonon system.
   In Fig.  \ref{Fig9}  we compare these values  for ZnO and anatase  (the factor $\hbar\omega_0$  is omitted).
 They  are given    for  the  excess energy of 0.3 eV, that is   the middle energy  between   the band   gap  and  the edge  of
 the solar  light, 4 eV.   The $f(\omega)$  value  is in this region slightly  less for  anatase than for ZnO,  but  the $\Gamma$   value
  for anatase is much higher.  Evidently, this is associated  with much higher density of states in the conduction band of anatase.
   Therefore  the rate  of  energy  loss in ZnO  appears to be about  5 times  less  than in anatase.
This is the factor  favoring  the higher  photocatalytic activity  of  ZnO.     Since  this  ratio  depends  on the states of only  conduction band
and on the electron-phonon coupling  of only these states,  one can expect  that  such ratio  is valid also  for the ZnO and  anatase  doped  with
elements producing  states  inside the band gap.    So  we may expect   that  the doped  ZnO  can   probably be  more  promising  material  for photocatalytic
applications than the doped anatase.   The photocatalytic  properties  of the doped ZnO have  been studied  much less than those of the doped  TiO$_2$.
Nevertheless, there are   few papers  containing the   comparison  of   activity  of ZnO and anatase  in the  reactions  of photo-decolarization
of red dye  \cite{Asl11,Kans09}.  It  has been shown in \cite{Asl11}  that the ZnO    catalyst composed of nano-particles   has  the activity   equal  to that of the analogous  anatase  catalyst,
 in  spite of  that  the  surface  area of  the anatase  catalyst  was more   extensive.  The  comparison of the  effectiveness of  ZnO- and TiO$_2$-based  photo-catalysts
in degradation  of dyes  made  by the authors of  the  Ref. \cite{Kans09}   demonstrated a 20 - 30 \% superiority  of  the ZnO-based  catalysts.
There are  also  examples of  the doped ZnO-based  catalysts  that have high  activity in visible light   \cite{Kana07,Shap09}.



\begin{thebibliography}{10}
\providecommand*{\bibinfo}[2]{#2}
\providecommand*{\eprint}[1]{#1}
\providecommand*{\url}[1]{#1}
\bibitem{Carp04}
\bibinfo{author}{O.~Carp}, \bibinfo{author}{C.~Huisman}, and
  \bibinfo{author}{A.~Reller}, \bibinfo{journal}{Progress in Solid State
  Chemistry} \bibinfo{volume}{\textbf{32}}, \bibinfo{pages}{33}
  (\bibinfo{date}{2004}).
\bibitem{Lu08}
\bibinfo{author}{H.~Lu}, \bibinfo{author}{H.~Li}, \bibinfo{author}{L.~Liao},
  \bibinfo{author}{Y.~Tian}, \bibinfo{author}{M.~Shuai}, \bibinfo{author}{J.~C.
  Li}, \bibinfo{author}{M.~Hu}, \bibinfo{author}{Q.~Fu}, and
  \bibinfo{author}{B.~Zhu}, \bibinfo{journal}{Nanotechnology}
  \bibinfo{volume}{\textbf{19}}, \bibinfo{pages}{045605}
  (\bibinfo{date}{2008}).
\bibitem{Ni07}
\bibinfo{author}{Y.~Ni}, \bibinfo{author}{X.~Cao}, \bibinfo{author}{G.~Wu},
  \bibinfo{author}{G.~Hu}, \bibinfo{author}{Z.~Yang}, and
  \bibinfo{author}{X.~Wei}, \bibinfo{journal}{Nanotechnology}
  \bibinfo{volume}{\textbf{18}}, \bibinfo{pages}{155603}
  (\bibinfo{date}{2007}).
\bibitem{Gird09}
\bibinfo{author}{O.~Girdasova}, \bibinfo{author}{V.~N. Krasilnikov},
  \bibinfo{author}{L.~I. Buldakova}, \bibinfo{author}{M.~Iu.Yanchenko}, and
  \bibinfo{author}{O.~V. Koriakova}, \bibinfo{journal}{Izv. Ross. Akad. Nauk,
  Ser. Fiz.} \bibinfo{volume}{\textbf{73}}, \bibinfo{pages}{1176}
  (\bibinfo{date}{2009}).
\bibitem{Hoff95}
\bibinfo{author}{M.~R. Hoffmann}, \bibinfo{author}{S.~Martin},
  \bibinfo{author}{W.~Choi}, and \bibinfo{author}{D.~Bahnemannt},
  \bibinfo{journal}{Chem. Rev.} \bibinfo{volume}{\textbf{95}},
  \bibinfo{pages}{69} (\bibinfo{date}{1995}).
\bibitem{Kim04}
\bibinfo{author}{S.~Kim}, \bibinfo{author}{W.-D. Kim},
  \bibinfo{author}{K.~Kim}, \bibinfo{author}{C.~Hwang}, and
  \bibinfo{author}{J.~Jeong}, \bibinfo{journal}{Appl. Phys. Lett.}
  \bibinfo{volume}{\textbf{85}}, \bibinfo{pages}{4112} (\bibinfo{date}{2004}).
\bibitem{Mai04}
\bibinfo{author}{C.~Maiti}, \bibinfo{author}{S.~Samanta},
  \bibinfo{author}{G.~Dalapati}, \bibinfo{author}{S.~Nandi}, and
  \bibinfo{author}{S.~Chatterjee}, \bibinfo{journal}{Microelectron. Eng.}
  \bibinfo{volume}{\textbf{72}}, \bibinfo{pages}{253} (\bibinfo{date}{2004}).
\bibitem{Lane04}
\bibinfo{author}{M.~Lane}, \bibinfo{author}{C.~Murray}, and
  \bibinfo{author}{F.~McFeely}, \bibinfo{journal}{Appl. Phys. Lett.}
  \bibinfo{volume}{\textbf{85}}, \bibinfo{pages}{4112} (\bibinfo{date}{2004}).
\bibitem{Yama99}
\bibinfo{author}{A.~Yamamoto}, \bibinfo{author}{T.~Kido},
  \bibinfo{author}{T.~Goto}, \bibinfo{author}{Y.~Chen},
  \bibinfo{author}{T.~Yao}, and \bibinfo{author}{A.~Kasuya},
  \bibinfo{journal}{Appl. Phys. Lett.} \bibinfo{volume}{\textbf{75}},
  \bibinfo{pages}{469} (\bibinfo{date}{1999}).
\bibitem{Hend07}
\bibinfo{author}{E.~Hendry}, \bibinfo{author}{M.~Koeberg}, and
  \bibinfo{author}{M.~Bonn}, \bibinfo{journal}{Phys. Rev. B}
  \bibinfo{volume}{\textbf{76}}, \bibinfo{pages}{045214}
  (\bibinfo{date}{2007}).
\bibitem{Wen07}
\bibinfo{author}{X.~Wen}, \bibinfo{author}{J.~Davis},
  \bibinfo{author}{D.~McDonald}, \bibinfo{author}{L.~Dao},
  \bibinfo{author}{P.~Hannaford}, \bibinfo{author}{V.~Coleman},
  \bibinfo{author}{H.~Tan}, \bibinfo{author}{C.~Jagadish},
  \bibinfo{author}{K.~Koike}, \bibinfo{author}{S.~Sasa}, \emph{et~al.},
  \bibinfo{journal}{Nanotechnolog} \bibinfo{volume}{\textbf{18}},
  \bibinfo{pages}{315403} (\bibinfo{date}{2007}).
\bibitem{Sun05}
\bibinfo{author}{C.-K. Sun}, \bibinfo{author}{S.-Z. Sun},
  \bibinfo{author}{K.-H. Lin}, \bibinfo{author}{K.~Y.-J. Zhang},
  \bibinfo{author}{H.-L. Liu}, \bibinfo{author}{S.-C. Liu}, and
  \bibinfo{author}{J.-J. Wu}, \bibinfo{journal}{Appl. Phys. Lett.}
  \bibinfo{volume}{\textbf{87}}, \bibinfo{pages}{023106}
  (\bibinfo{date}{2005}).
\bibitem{Wata00}
\bibinfo{author}{M.~Watanabe}, \bibinfo{author}{S.~Sasaki}, and
  \bibinfo{author}{T.~Hayashi}, \bibinfo{journal}{J. Lumin.}
  \bibinfo{volume}{\textbf{87–89}}, \bibinfo{pages}{1234}
  (\bibinfo{date}{2000}).
\bibitem{Wata01}
\bibinfo{author}{M.~Watanabe}, \bibinfo{author}{T.~Hayashi},
  \bibinfo{author}{H.~Yagasaki}, and \bibinfo{author}{S.~Sasaki},
  \bibinfo{journal}{Int. J. Mod. Phys. B} \bibinfo{volume}{\textbf{15}},
  \bibinfo{pages}{3997} (\bibinfo{date}{2001}).
\bibitem{Seki04}
\bibinfo{author}{T.~Sekiya}, \bibinfo{author}{M.~Tasaki},
  \bibinfo{author}{K.~Wakabayashi}, and \bibinfo{author}{S.~Kurita},
  \bibinfo{journal}{Journal of Luminesc.} \bibinfo{volume}{\textbf{108}},
  \bibinfo{pages}{69} (\bibinfo{date}{2004}).
\bibitem{Hend04}
\bibinfo{author}{E.~Hendry}, \bibinfo{author}{F.~Wang},
  \bibinfo{author}{J.~Shan}, \bibinfo{author}{T.~F. Heinz}, and
  \bibinfo{author}{M.~Bonn}, \bibinfo{journal}{Phys. Rev. B}
  \bibinfo{volume}{\textbf{69}}, \bibinfo{pages}{081101}
  (\bibinfo{date}{2004}).
\bibitem{Zhuk02}
\bibinfo{author}{V.~P. Zhukov}, \bibinfo{author}{E.~Chulkov}, and
  \bibinfo{author}{P.~Echenique}, \bibinfo{journal}{Physical Review B}
  \bibinfo{volume}{\textbf{65}}, \bibinfo{pages}{115116}
  (\bibinfo{date}{2002}).
\bibitem{Zhuk04}
\bibinfo{author}{V.~P. Zhukov}, \bibinfo{author}{E.~Chulkov}, and
  \bibinfo{author}{P.~Echenique}, \bibinfo{journal}{Physical Review Letters}
  \bibinfo{volume}{\textbf{93}}(9), \bibinfo{pages}{096401}
  (\bibinfo{date}{2004}).
\bibitem{Knor02a}
\bibinfo{author}{R.~Knorren}, \bibinfo{author}{G.~Bouzerar}, and
  \bibinfo{author}{K.~Bennemann}, \bibinfo{journal}{J. Phys.: Condens. Matter}
  \bibinfo{volume}{\textbf{14}}, \bibinfo{pages}{R739} (\bibinfo{date}{2002}).
\bibitem{Baro01}
\bibinfo{author}{S.~Baroni}, \bibinfo{author}{S.~de~Gironcoli}, and
  \bibinfo{author}{A.~D. Corso}, \bibinfo{journal}{Rev. of Modern Physics}
  \bibinfo{volume}{\textbf{73}}, \bibinfo{pages}{515} (\bibinfo{date}{2001}).
\bibitem{Zein84}
\bibinfo{author}{E.~Zein}, \bibinfo{journal}{Sov. Phys.—Solid State}
  \bibinfo{volume}{\textbf{26}}, \bibinfo{pages}{1825} (\bibinfo{date}{1984}).
\bibitem{Zhuk10b}
\bibinfo{author}{V.~Zhukov}, \bibinfo{author}{P.~Echenique}, and
  \bibinfo{author}{E.~Chulkov}, \bibinfo{journal}{Phys. Rev. B}
  \bibinfo{volume}{\textbf{82}}, \bibinfo{pages}{094302}
  (\bibinfo{date}{2010}).
\bibitem{Zhuk10a}
\bibinfo{author}{V.~Zhukov} and \bibinfo{author}{E.~Chulkov},
  \bibinfo{journal}{J. Phys.: Condens. Matter} \bibinfo{volume}{\textbf{22}},
  \bibinfo{pages}{435802} (\bibinfo{date}{2010}).
\bibitem{Tyut11}
\bibinfo{author}{V.G.Tyuterev}, \bibinfo{author}{S.~Obukhov},
  \bibinfo{author}{N.~Vast}, and \bibinfo{author}{J.~Sjakste},
  \bibinfo{journal}{Phys. Rev. B} \bibinfo{volume}{\textbf{84}},
  \bibinfo{pages}{035201} (\bibinfo{date}{2011}).
\bibitem{DasS88}
\bibinfo{author}{S.~D. Sarma}, \bibinfo{author}{J.~K. Jain}, and
  \bibinfo{author}{R.~Jalabert}, \bibinfo{journal}{Physical Review B}
  \bibinfo{volume}{\textbf{37}}, \bibinfo{pages}{6290} (\bibinfo{date}{1988}).
\bibitem{Komi01}
\bibinfo{author}{S.~Komirenko}, \bibinfo{author}{K.~Kim},
  \bibinfo{author}{M.~Stroscio}, and \bibinfo{author}{M.~Dutta},
  \bibinfo{journal}{J. Phys.: Condens. Matter} \bibinfo{volume}{\textbf{13}},
  \bibinfo{pages}{6233} (\bibinfo{date}{2001}).
\bibitem{Jang07}
\bibinfo{author}{D.-J. Jang}, \bibinfo{author}{G.-T. Lin},
  \bibinfo{author}{C.-L. Wu}, \bibinfo{author}{C.-L. Hsiao}, and
  \bibinfo{author}{L.~W. Tu}, \bibinfo{journal}{Appl. Phys. Lett.}
  \bibinfo{volume}{\textbf{91}}, \bibinfo{pages}{092108}
  (\bibinfo{date}{2007}).
\bibitem{Tao93}
\bibinfo{author}{Z.~Tao}, \bibinfo{author}{C.~S. Ting}, and
  \bibinfo{author}{M.~Singh}, \bibinfo{journal}{Phys. Rev. Lett.}
  \bibinfo{volume}{\textbf{70}}, \bibinfo{pages}{2467}.
\bibitem{Maha90}
\bibinfo{author}{G.~Mahan}, \bibinfo{title}{\emph{Many-particle physics}}
  (\bibinfo{publisher}{Plenum Press}, New York, \bibinfo{year}{1990}).
\bibitem{Grim}
\bibinfo{author}{G.~Grimvall}, \bibinfo{title}{\emph{The Electron-Phonon
  Interactions in Metals}} (\bibinfo{publisher}{North-Holland}, Amsterdam,
  \bibinfo{year}{1981}).
\bibitem{Ande87}
\bibinfo{author}{O.~Andersen}, \bibinfo{author}{O.~Jepsen}, and
  \bibinfo{author}{M.~Sob}, in \bibinfo{editors}{M.~Yussouff}, ed.,
  \emph{Electronic band structure and its applications}
  (\bibinfo{publisher}{Springer}, \bibinfo{year}{1987}), \bibinfo{volume}{vol.
  283 of \emph{Lecture Notes in Physics}}.
\bibitem{Loud83}
\bibinfo{author}{R.~Loudon}, \bibinfo{title}{\emph{The quantum theory of
  light}} (\bibinfo{publisher}{Oxford University Press Inc.}, New York,
  \bibinfo{year}{1983}).
\bibitem{Zain10a}
\bibinfo{author}{V.~Zainullina}, \bibinfo{author}{M.~Korotin}, and
  \bibinfo{author}{V.~Zhukov}, \bibinfo{journal}{Physica B}
  \bibinfo{volume}{\textbf{405}}, \bibinfo{pages}{2110} (\bibinfo{date}{2010}).
\bibitem{Usud02}
\bibinfo{author}{M.~Usuda}, \bibinfo{author}{N.~Hamada},
  \bibinfo{author}{T.~Kotani}, and \bibinfo{author}{M.~van Schilfgaarde},
  \bibinfo{journal}{Phys. Rev. B} \bibinfo{volume}{\textbf{66}},
  \bibinfo{pages}{125101} (\bibinfo{date}{2002}).
\bibitem{Emor10}
\bibinfo{author}{M.~Emori}, \bibinfo{author}{M.~Sugita},
  \bibinfo{author}{H.~Sakama}, and \bibinfo{author}{K.~Ozawa},
  \bibinfo{journal}{Photon Factory Activity Report 2009 \# 27 Part B}
  \bibinfo{pages}{p.~98} (\bibinfo{date}{2010}).
\bibitem{Cui08}
\bibinfo{author}{Y.~Cui}, \bibinfo{author}{H.~Du}, and
  \bibinfo{author}{L.~Wen}, \bibinfo{journal}{J. Mater. Sci. Technol.}
  \bibinfo{volume}{\textbf{24}}, \bibinfo{pages}{675} (\bibinfo{date}{2008}).
\bibitem{Asl11}
\bibinfo{author}{S.~K. Asl}, \bibinfo{author}{S.~K. Sadrnezhaad},
  \bibinfo{author}{M.~K. Rad}, and \bibinfo{author}{D.~\"{E}uner},
  \bibinfo{journal}{Turk J. Chem} \bibinfo{volume}{\textbf{35}},
  \bibinfo{pages}{1} (\bibinfo{date}{2011}).
\bibitem{Kans09}
\bibinfo{author}{S.~K. Kansal}, \bibinfo{author}{N.~Kaur}, and
  \bibinfo{author}{S.~Singh}, \bibinfo{journal}{Nanoscale Res. Lett.}
  \bibinfo{volume}{\textbf{4}}, \bibinfo{pages}{709} (\bibinfo{date}{2009}).
\bibitem{Shap09}
\bibinfo{author}{A.~Shaporev}, Master's thesis, Institute of General and
  Inorganic chemistry, Leninski prospect, 31, 119991 Moscow, Russian Federation
  (\bibinfo{date}{2009}).
\bibitem{Kana07}
\bibinfo{author}{K.~Kanade}, \bibinfo{author}{B.~Kale}, \bibinfo{author}{J.-O.
  Baeg}, \bibinfo{author}{S.~M. Lee}, \bibinfo{author}{C.~W. Lee},
  \bibinfo{author}{S.-J. Moon}, and \bibinfo{author}{H.~Chang},
  \bibinfo{journal}{Materials Chemistry and Physics}
  \bibinfo{volume}{\textbf{102}}, \bibinfo{pages}{98} (\bibinfo{date}{2007}).
\end{thebibliography}

\end{document}